# An Optimized Quantum Maximum or Minimum Searching Algorithm and its Circuits

Yanhu Chen[1], Shijie Wei[2,3], Xiong Gao[1], Cen Wang[1], Jian Wu[1], Hongxiang Guo[1,*]

[1]*Institute of Information Photonics and Optical Communications, Beijing University of Posts and Telecommunications, Beijing 100876, China*

[2]*Beijing Academy of Quantum Information Sciences，Beijing 100193, China*

[3]*State Key Laboratory of Low-Dimensional Quantum Physics and Department of Physics, Tsinghua University, Beijing 100084, China*

**Abstract:** Finding a maximum or minimum is a fundamental building block in many mathematical models. Compared with classical algorithms, Durr, Hoyer's quantum algorithm (DHA) achieves quadratic speed. However, its key step, the quantum exponential searching algorithm (QESA), which is based on Grover algorithm, is not a sure-success algorithm. Meanwhile, quantum circuits encounter the gate decomposition problem due to variation of the scale of data. In this paper, we propose an optimized quantum algorithm for searching maximum and minimum, based on DHA and the optimal quantum exact search algorithm. Furthermore, we provide the corresponding quantum circuits, together with three equivalent simplifications. In circumstances when we can exactly estimate the ratio of the number of solutions $M$ and the searched space $N$, our method can improve the successful probability close to 100%. Furthermore, compared with DHA, our algorithm shows an advantage in complexity with large databases and in the gate complexity of constructing oracles. Experiments have been executed on an IBM superconducting processor with two qubits, and a practical problem of finding the minimum from Titanic passengers' age was numerically simulated. Both showed that our optimized maximum or minimum performs more efficiently compared with DHA. Our algorithm can serve as an important subroutine in various quantum algorithms which involves searching maximum or minimum.

# 1. Introduction

In the era of data explosion, computing devices have to process more and more data with higher speed and better efficiency. Cisco forecasts that 77 EB data traffic will be generated per month by 2022[1]. But the computing power of classical computers tends to reach its upper limit. It is urgent to find a new way to process large-scale data.

By exploiting properties of quantum mechanics, researchers have discovered some remarkable quantum algorithms to accelerate a range of algorithms in quantum computers [2]. In 1985, Deutsch's algorithm could decide whether a function is constant or balanced by only one call of the function, but any classical algorithm requires two calls of the function [3,4]. In 1994, Shor's algorithm for factoring achieved exponential speed increase [5]. In 1996, a fast quantum search algorithm was discovered by Grover, which solves the searching problem by using approximately $\sqrt{N}$ operations rather than approximately $N$ operations in the classical algorithm[6-8]. Diao pointed out, a strictly exact search is possible only if the ratio of solutions $M$ to the database size $N$ is $1/4$[9]. Especially, the highest failure rate is 50% when $M/N = 1/2$. Various generalized and modified versions of Grover algorithm, including the phase matching methods, have been explored with the view of improving the efficiency of Grover algorithm [10-16]. Among them, the Grover-Long algorithm has one adjustable phase that finds the target with zero failure rate for any database [16], with exactly the same number of iterations as that of the standard Grover algorithm. Ref. [16] actually provides a series of exact quantum search algorithms, each with an iteration number $J \geq J_0$, where $J_0$ is the number of iterations in the Grover algorithm. The exact quantum search algorithms are usually called Long's algorithm [20,21], and the optimized one with iteration number is called Grover-Long algorithm [21], which has been shown by Toyama et al to be exactly optima. Meanwhile, many quantum algorithms based on Grover algorithm for various applications [17-23], such as finding maximum/minimum [22-23], were proposed.

Classically, searching the maximum/minimum problem requires approximately $N$ operations, but its quantum counterpart, which was proposed by Durr, Hoyer based on the quantum exponential searching algorithm (QESA) [17,22], achieves quadratic speedup. When the number of solutions is unknown, QESA reduces its failure rate at the expense of repeatedly performing Grover's algorithm with different number of iterations. However, since Grover algorithm is not a sure-success algorithm, QESA is not optimal. Besides, the preparation of the initial state in the repetition approach also takes time. Therefore, it is possible and desirable to optimize the Durr, Hoyer algorithm. In this paper, we propose an optimized quantum maximum or minimum searching algorithm (QUMMSA), which provides two improvements. First, it removes the theoretical failure rate of Grover algorithm by replacing QESA with Grover-Long algorithm. By using the minimum sample size to estimate the ratio of the number of solutions and the database size $\frac{M}{N}$, we can obtain exact parameters of Grover-Long algorithm. Second, it replaces the interruption condition of DHA with a constant $c$, which is independent of the database size. The failure rate decreases exponentially as $c$ increases linearly.

Demonstrating quantum algorithms in a real quantum information processing device is very important in the development of quantum algorithms. For example, Grover algorithm has been demonstrated on a variety of physical platforms, such as NMR systems[24-26], superconducting processors[27], trapped atomic ions[28-30] and photonics[31-32]. To implement quantum algorithms, one must design the quantum circuits. However, it is hard to decompose a high dimensional unitary matrix into elementary quantum gates to solve issues of different scale. In this paper, we propose a general method to design quantum circuits for the QUMMSA and it can be integrated into a general-purpose quantum software. Due to the characteristics of the problem, the circuit design can further reduce the complexity of gates in constructing the oracles. The designed circuits are easier to implement on any general-purpose quantum computer.

The remainder of this paper is organized as follows. In Section 2, we presented QUMMSA based on Grover-Long algorithm. In Section 3, we gave the general quantum circuits, and three equivalent simplified principles. In Section 4, an experiment implemented in an IBM superconducting processor, and a numerical simulation of a 6-qubits system to solve a specific problem were presented. They showed that QUMMSA is indeed more efficient. In Section 5, we analyzed the failure rate of the QUMMSA, and proposed two methods to further reduce the failure rate in big data scenarios. The complexity of the two algorithms are compared. In Section 6, we drew a conclusion and gave an outlook of QUMMSA in future applications.

## 2. The QUMMSA

**Problem:** Let $D$ be an unsorted database with $N$ items. The problem is to find the maximum or minimum from $D$. For convenience, we only present the minimum searching algorithm as an example. The maximum searching algorithm can be achieved similarly.

**Core idea:** Exploiting the Grover-Long algorithm, we can find $M(M \geq 1)$ solutions from the unsorted database with $N$ items. Here, a random value $d_0$ is taken as a reference value. If the search algorithm gives a result $d_1$, which is less than or equals to $d_0$, it will run successfully. Note that there are $M$ results that satisfy the search condition and $d_1$ is one of them. Then, let $d_1$ replace $d_0$ and repeat the above steps until $M = 1$. Since the $M$ solutions are given with equal probability after Grover-Long algorithm, the number of solutions will be reduced by half on average, after one main loop. Therefore, the mathematical expectation of main loops to find the minimum is $\log_2 N$, in theory.

**Hypotheses:** To simplify the problem, our hypotheses are as follows.

(1) Each data value is represented by a binary string and is stored in an orthonormal basis state of $|\Psi\rangle$, where $|\Psi\rangle$ is the initial state. Therefore, the data value lies in interval $[0, 2^n - 1]$, where $n$ is the number of qubits.

(2) There is a one-to-one mapping between a data value and its index. The index may be a person's name or other non-numeric data.

(3) Each data value is an integer.

(4) Each data value is distinct.

(5) Preparing an initial state takes $\log_2(N)$ steps. Performing an oracle takes one step. Others are not counted.

(6) One orthonormal basis state stores a data value, the amplitude is $1/\sqrt{N}$ ($2^{n-1} < N \leq 2^n$). The amplitude will be 0 if no data value is stored.

In summary, hypotheses (1-2) define the quantum data type which is similar to the data type of classical computers. For example, uint8 is a classical data type which means 8 bits are used to store an unsigned integer. The data type limits the range of data values. Without loss of generality, we make the above hypotheses (3-5) as used in Ref [22]. In original Grover algorithm, the initial state is a uniform superposition state, which doesn't apply to the QUMMSA. Hypothesis (6) indicates that not all orthonormal basis states are stored with data values.

**Pseudocode:** Here we provide the pseudocode of QUMMSA, as shown in Algorithm 1.

**Algorithm 1:** Quantum algorithm for finding the minimum

| | | |
|---|---|---|
| 1 | **Input**: An unsorted database *D*. | |

| | |
|---|---|
| | A random value $d_0$ which is chosen from $D$. |
| 2 | **Output**: The minimum of the unsorted database. |
| 3 | **function** $QMIN(D, d_0)$ |
| 4 | $d_1 = +\infty$; |
| 5 | Set a positive integer $c$; |
| 6 | **for** (*i=0*; *i<c*; *i=i+1*) |
| 7 | Design an oracle according to $d_0$, i.e., the oracle can mark all data values which is less than or equal to $d_0$; |
| 8 | **while** $d_1 > d_0$ |
| 9 | Map $D$ to an initial state $|\Psi\rangle$. |
| 10 | Apply Grover-Long algorithm on the initial state $|\Psi\rangle$; |
| 11 | Measure the quantum register and assign the result to $d_1$; |
| 12 | **end** |
| 13 | **if** $d_1 < d_0$ |
| 14 | *i=0*; |
| 15 | **end** |
| 16 | $d_0 = d_1$; |
| 17 | **end** |
| 18 | **return** the minimum value $d_0$. |

Compared with DHA, the QUMMSA provides two improvements. First, removing the theoretical failure rate of Grover algorithm by replacing QESA with Grover-Long algorithm. Through the sample estimation, we can obtain close to 100% accurate parameters of Grover-Long algorithm, even if $M$ and $N$ are unknown. Second, it replaces the interrupt condition of DHA with a constant $c$ which is independent of the database size. In the worst case, QUMMSA has a $1 - 1/2^c$ possibility to find the minimum. In Section 5.1, we will discuss the performance of QUMMSA, in detail. Since Grover-Long algorithm is a key step of QUMMSA, we will describe a general design method for quantum circuits of Grover-Long algorithm.

## 3. General circuits and three equivalent simplified principles

### 3.1 Review of Grover-Long algorithm

The initial state can be prepared by $W$ operator, which can be described as formula(1):

$$|\Psi\rangle = W|0^{\otimes n}\rangle = \frac{1}{\sqrt{N}}\sum_{i=0}^{N-1}|i\rangle = \sqrt{\frac{M}{N}}|\Psi_{good}\rangle + \sqrt{\frac{N-M}{N}}|\Psi_{bad}\rangle \quad (1)$$

Where $|\Psi_{good}\rangle$ stores solutions which we want to find and $|\Psi_{bad}\rangle$ stores other values; $N$ is the database size; $M$ is the number of solutions. Especially, when $N = 2^n$, the initial state is a uniform superposition state, the $W$ operator becomes $H^{\otimes n}$, where $H$ is the Walsh-Hadamard transformation; $n$ is the number of qubits.

One Grover iteration can be divided into four operators.

$$G = -WI_0W^{-1}O \quad (2)$$

Where $O$ is an oracle which performs a phase inversion on $|\Psi_{good}\rangle$; $I_0$ is a conditional phase shift operator which performs a phase inversion on $|0\rangle$.

Grover-Long algorithm is done by replacing the phase inversion with an adjustable angle $\phi$ phase rotation. The rotation angle is given as:

$$\phi = 2\arcsin\left(\frac{\sin\frac{\pi}{4J+2}}{\sin\beta}\right) \quad (3)$$

Where $\sin\beta = \sqrt{\frac{M}{N}}$. Upon measurement in $J$-th iteration, one of marked states is obtained with zero failure rate.

$$J \geq \text{floor}\left(\frac{\frac{\pi}{2}-\beta}{\beta}\right) + 1 \quad (4)$$

By utilizing the number of solutions $M$ and the database size $N$, we can calculate the exact value of $\beta$, $\phi$, $J$. Grover-Long algorithm will find a solution with zero failure rate.

## 3.2 The design of $I_0$ operator

Since only $|0\rangle$ receives a rotation phase, the operator of $I_0$ can be described as a diagonal matrix, as shown in formula (5).

$$I_0 = e^{i\phi}|0\rangle\langle 0| + \sum_{\tau=1}^{2^n-1}|\tau\rangle\langle\tau| = \text{diag}[e^{i\phi}, 1, \ldots, 1]_{2^n} \tag{5}$$

The first element of $I_0$ is always $e^{i\phi}$, other elements are 1. $n$ is the number of qubits. $I_0$ can be converted to the quantum circuit, as shown in Fig 1.

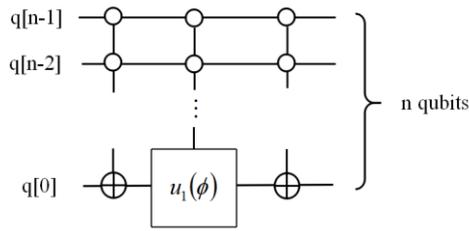

**Fig 1.** The general circuit for $I_0$ operator, where q[0] denotes the lowest qubit, q[n-1] denotes the highest qubit.

## 3.3 The design of oracle $O$

Oracle can recognize the solutions of a searching problem. If one orthonormal basis state is one of solutions, it will receive a rotation phase. Here, we elaborate on the construction of oracle from two parts: the searching problem has a unique solution or multiple solutions.

Firstly, there is only one solution in the searching problem. Namely, the oracle can be described as a diagonal matrix that has only one $e^{i\phi}$, as shown in formula (6).

$$O = e^{i\phi}|v\rangle\langle v| + \sum_{\tau=0,\tau\neq v}^{2^n-1}|\tau\rangle\langle\tau| \tag{6}$$

Where $v$ is the position of $e^{i\phi}$ in the diagonal matrix.

The position $v$ of $e^{i\phi}$ is divided into two cases. If $v$ is odd, the $u_1(\phi)$ gate will be applied to q[0], where $u_1(\phi) = \text{diag}[1, e^{i\phi}]$. If $v$ is even, X, $u_1(\phi)$, X gates will be applied to q[0]. As shown in Fig 2 and Fig3, q[0] is a target qubit and other qubits are control qubits. Oracle operators can be converted to quantum circuits, where the white dot of the $j$-th($1 \leq j \leq n-1$) line denotes that the operator is applied to q[0] when the $j$-th qubit is set to $|0\rangle$; the black dot denotes that the operator is applied to q[0] when the qubit is set to $|1\rangle$. $v$ can be expressed by $v = 1 + \sum_{j=1}^{n-1} g(j) \times 2^j$, when the $u_1(\phi)$ gate is applied to q[0]. Another case, $v = \sum_{j=1}^{n-1} g(j) \times 2^j$, when X,$u_1(\phi)$,X gates are applied to q[0]. Note that $g(j)$ is a bool function which denotes the $j$-th qubit is 0 or 1. Through the above steps, the oracle can mark any quantum state by changing the position of $e^{i\phi}$, when the searching problem has a unique solution, where $n$ is any positive integer.

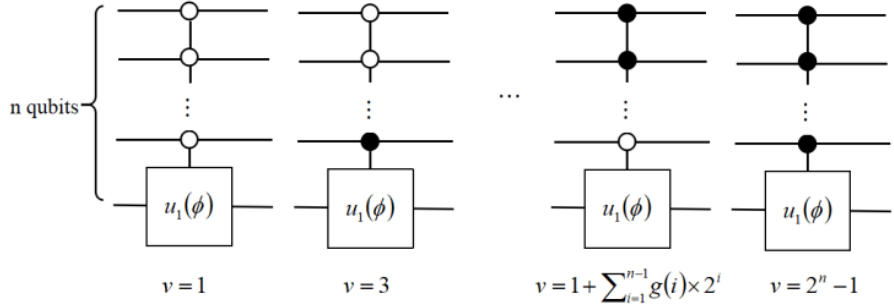

**Fig 2.** General circuits for different oracles which mark an odd state.

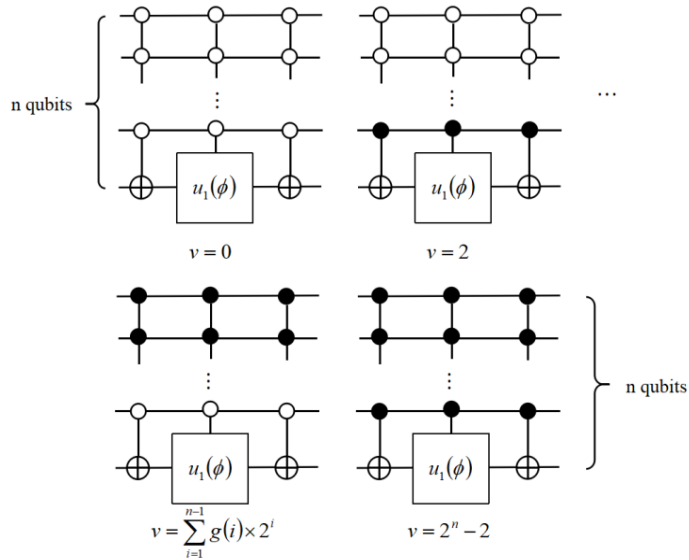

**Fig 3.** General circuits for different oracles which mark an even state.

Secondly, we should discuss that the oracle can mark $M$ $(0 < M \leq 2^n)$ quantum states. Namely, the number of solutions is $M$. The oracle can be described as a diagonal matrix.

$$O = e^{i\phi} \sum_{\tau=1}^{M} |v_\tau\rangle\langle v_\tau| + \sum_{\tau=0, \tau \notin V}^{2^n-1} |\tau\rangle\langle\tau| \tag{7}$$

Where the number of $e^{i\phi}$ is $M$ and the set of $e^{i\phi}$ position is $V = \{v_1, v_2, ..., v_M\}$. The oracle marking multiple quantum states can be composed of many oracles that mark one quantum state.

For example, if oracle marks two quantum states which $V = \{0,1\}$. Then, $u_1(\phi)$ and $X,u_1(\phi),X$ gates are applied to q[0], other control qubits follow the previous rules, as shown in Fig 4.

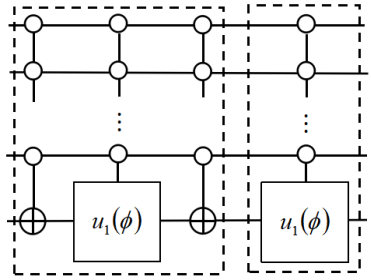

**Fig 4.** A general circuit for marking states $|0\rangle$ and $|1\rangle$.

## 3.4 Three equivalent simplified principles

If there are $2^m$ $(1 \leq m \leq n-1)$ solutions, the oracle will become very complex. Besides, it's difficult to execute too many entanglement gates on current quantum computers. Here, we proposed three principles to simplify circuit construction.

Firstly, it is well-known that an $n$-qubit controlled phase gate can be approximately decomposed into $2^{n-1}$ two-qubit controlled phase gates [33]. Thus if oracle marks $2^m$ states, $2^{n+m-1}$ two-qubit controlled gates will be performed. If the searching problem is finding the minimum value, the oracle will mark all values less than or equal to $d_0$. Under such conditions, we proposed the first principle which only uses $2^{n-m-1}$ two-qubit controlled gates to mark $2^m$ states. The schematic diagram is shown in Fig 5.

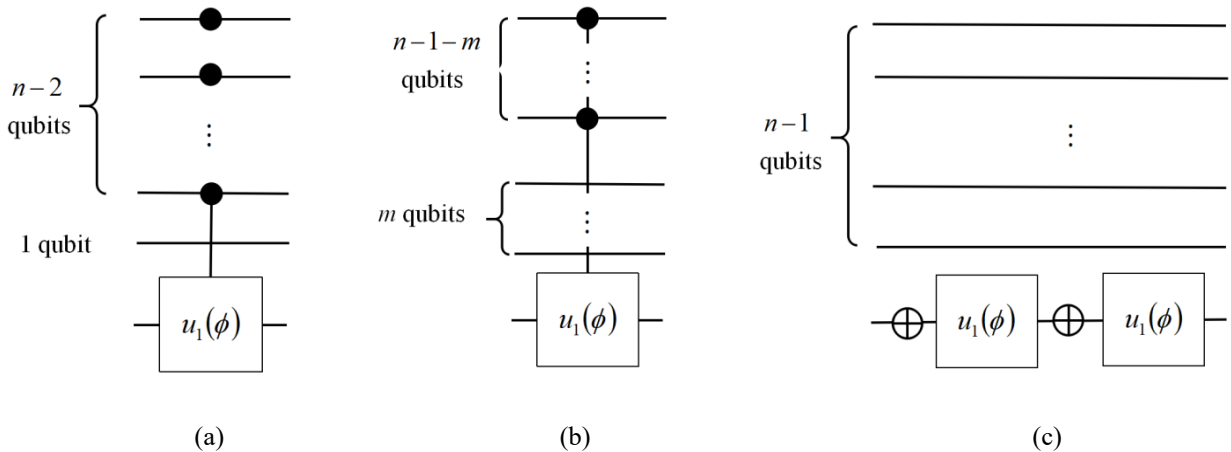

**Fig 5.** A schematic diagram of the first equivalent simplified principle. (a) The circuit for marking two continuous odd states by an $(n-1)$-qubit controlled phase gate(it has $n-2$ control qubits and a single target qubit). (b) The circuit for marking $2^m$ continuous odd states. (c) The circuit for marking all states.

Secondly, since a multi-qubit controlled gate error is far more than a single-qubit gate error in most types of quantum computers, it's necessary to use the number of multi-qubit controlled gates as few as possible. We proposed the second principle for those oracles that cannot be simplified by the first principle, such as a single even state. The circuit has the same multi-qubit $CNOT$ gate on the pre- and post-controlled phase gate, such as $I_0$ operator. The multi-qubit $CNOT$ gate can be simplified to a $NOT$ gate as shown in Fig 6. Therefore, this oracle only uses an n-qubit controlled gate(it has $n-1$ control qubits and a single target qubit).

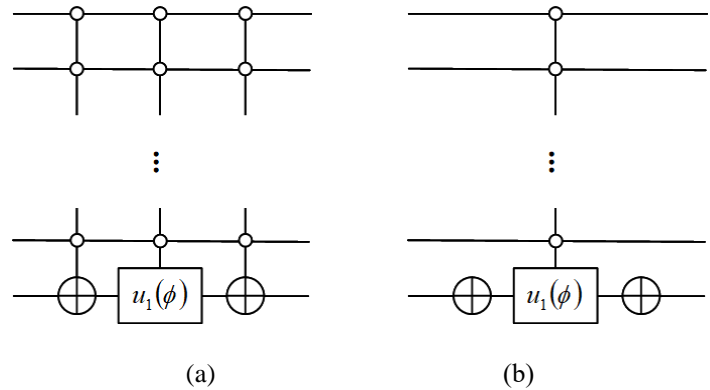

**Fig 6.** A schematic diagram of the second equivalent simplified principle. (a) The original circuit. (b) The simplified circuit.

Thirdly, to simplify the oracles marking even and odd states, we proposed the third principle, through elementary algebraic transformation [36]. After the above two simplified principles, the oracle will contain several circuits similar to Fig 7(a) which can be simplified as Fig 7(b).

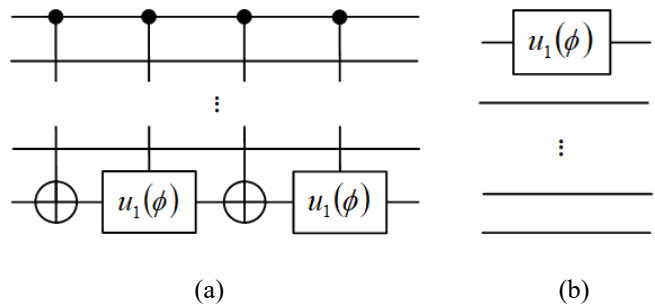

**Fig 7.** A schematic diagram of the third equivalent simplified principle. (a) The original circuit. (b) The simplified circuit.

# 4. Experiment and simulation

In this section, we compare the key step of DHA (with QESA) and QUMMSA (with Grover-Long) firstly by a 2-qubit experiment based on a superconducting processor. Besides, a 6-qubit numerical simulation was conducted to show how QUMMSA can efficiently solve a minimum finding problem based on a real data set (passenger age (excerpt) of the Titanic). Through the results of two demos, we report that comparing to QESA in DHA, QUMMSA possesses shorter circuit depth, less multi-qubit gates and lower failure rate by means of Grover-Long algorithm.

## 4.1 A 2-qubit contrast experiments

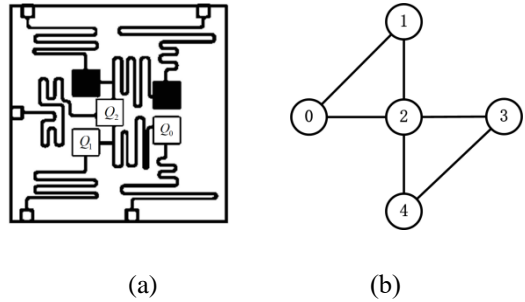

(a)          (b)

**Fig 8.** 5-qubit superconducting processor: (a)schematic; (b) topology.

The experimental device is IBMQ Yorktown which consists of five coupled superconducting transmons. Limited by the accuracy of the experimental device, two qubits $Q_0$, $Q_2$ were used for the experiment of Grover-Long algorithm. While, two work qubits $Q_1$, $Q_2$ and an ancilla qubit $Q_0$ were used for QESA. The schematic and topology of this processor are shown in Fig 8(a) and Fig 8(b) respectively. Two co-planar waveguide (CPW) resonators, acting as quantum buses, provide the device control and readout. Entanglement in IBM system is achieved via CNOT gates, which use cross-resonance [34,35]. Single qubit rotation gate with an arbitrary angle and CNOT are as primitive operators. The details of the device parameters are attached in Appendix A.

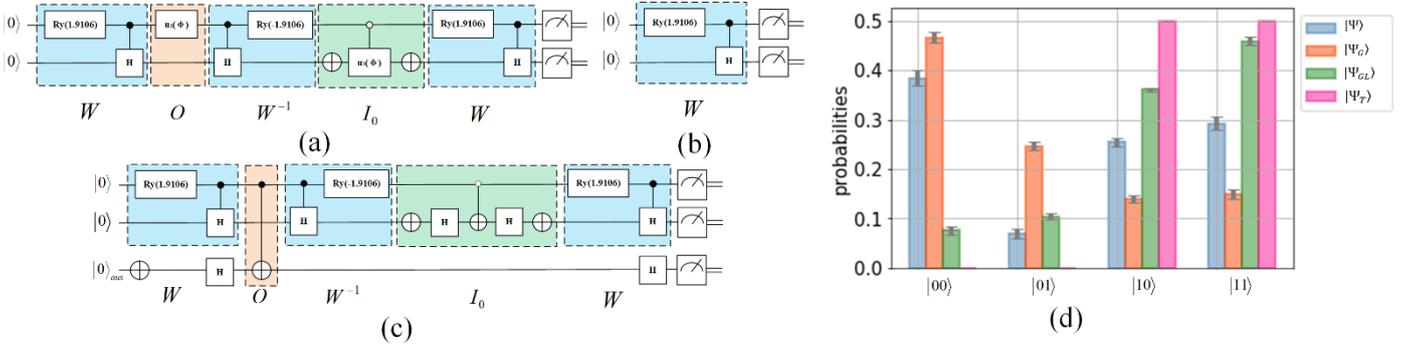

**Fig 9.** (a) A 2-qubit circuit of Grover-Long algorithm. (b) A 2-qubit circuit of the initial state preparation. (c) A 2-qubit and an ancilla qubit circuit of Grover algorithm [36]. (d) The experimental result and the theoretical result. Among them, $|\Psi\rangle$ is the initial state; $|\Psi_G\rangle$ is the state after applying Grover algorithm; $|\Psi_{GL}\rangle$ is the state after applying Grover-Long algorithm; $|\Psi_T\rangle$ is the ideal result.

In this demo, aiming at finding maximum, we set $N \in \{3,4\}$ as the database size, $M \in \{1,2,3\}$ as the number of solutions. Even the exact $M$ and $N$ are set in advance, in the experiment, but they are unknown in real data searching scenario, thus estimated $\widetilde{M}$ and $\widetilde{N}$ are used. Due to prior knowledge missing, we would suppose that each orthogonal basis state stores a value. Considering initial states, though combination, 12 kinds of circuits can be obtained for each algorithm.

A specific combination ($N = 3$, $M = 2$, formula (8) as the initial state) is taken for example to show details.

$$|\Psi\rangle = \frac{1}{\sqrt{3}}[1,0,1,1]^T \tag{8}$$

Given randomly select value from the database $d_0 = 2$, any state that $\geq d_0$ should be marked, namely $|10\rangle$, $|11\rangle$ in this case. Then after applying two algorithms, the measurement result $d_1$ could be obtained. If $d_1 \geq d_0$, the algorithm is thought to operate successfully.

(1) **The experiment of Grover-Long algorithm**

Two qubits $Q_0$, $Q_2$ which have the best performance in IBMQ Yorktown were used for Grover-Long algorithm. The estimated value of the database size was set as $\widetilde{N} = 2^n = 4$ and the estimated value of the number of solutions was set as $\widetilde{M} = 2^n - d_0 = 2$. Then the estimated $\tilde{\beta} = 0.7854$, $\tilde{J}=1$, and $\widetilde{\Phi} = \pi/2$ could be calculated by formula (3,4). After parameters estimation, Grover-Long algorithm was applied with $\tilde{J}$ iterations on the initial state $|\Psi\rangle$. The optimized circuit following the construction rules of Section 3 is shown in Fig 9(a). The $R_y(\theta)$ is defined as an operator of the rotation $\theta$ angle around the Y-axis.

$$R_y(\theta) = \begin{bmatrix} \cos\frac{\theta}{2} & -\sin\frac{\theta}{2} \\ \sin\frac{\theta}{2} & \cos\frac{\theta}{2} \end{bmatrix} \quad (9)$$

Owing to the error between the estimated value $\frac{\widetilde{M}}{\widetilde{N}}$ and the exact value $\frac{M}{N}$, the 2-qubit theoretical failure rate $\varepsilon_{GL}$ will exist and can be calculated by formula (10) which is the specific case of formula(C7).

$$\varepsilon_{GL} = \text{abs}\left(-e^{i\widetilde{\phi}}(e^{i\widetilde{\phi}}-1)\sqrt{\frac{M}{N}\times\left(1-\frac{M}{N}\right)}\times\sqrt{\frac{M}{N}}+\left[-e^{i\widetilde{\phi}}+(e^{i\widetilde{\phi}}-1)\times\frac{M}{N}\right]\times\left(1-\frac{M}{N}\right)^{\frac{1}{2}}\right)^2 \approx 0.037 \quad (10)$$

However, the experimental failure rate of Grover-Long algorithm $\varepsilon_{GL_E}$ is 0.180, because of the gate error and readout error of IBMQ.

(2) **The QESA experiment**

Two work qubits $Q_1$, $Q_2$ were used to operate QESA, and an ancilla qubit $Q_0$ initialized to $|1\rangle$ was added according to standard Grover algorithm requirement. In any $t$-th iterations of QESA, the number of Grover iteration $\gamma$ was randomly generated and rounded down from $[0,\lambda^{t-1})$, where $\lambda\in(1,\frac{4}{3}]$[17]. When $t=1$, $\gamma=0$, the theoretical failure rate $\varepsilon_{ESA}^{(1)}$ is described as formula(11) which is the specific case of formula(C9).

$$\varepsilon_{ESA}^{(1)} = 1 - \frac{M}{N} \quad (11)$$

If this measurement result is incorrect, the QESA will run the next iteration. When $t$-th iteration of QESA, the initial state is directly measured(the circuit is shown in Fig 9(b)) under the probability of $\lambda^{-t+1}$, while measurement after one Grover iteration is operated(the circuit is shown in Fig 9(c)) under the probability of $1-\lambda^{t-1}$. So the theoretical failure rate $\varepsilon_{ESA}^{(t)}$ can be described as formula(12).

$$\varepsilon_{ESA}^{(t)} = \varepsilon_{ESA}^{(t-1)} \times \left[\lambda^{-t+1}\left(1-\frac{M}{N}\right)+(1-\lambda^{-t+1})\cos^2\left(3\times\arcsin\sqrt{\frac{M}{N}}\right)\right] \quad (12)$$

Where, $\varepsilon_{ESA}^{(t)}$ is the failure rate after running the $t$-th iteration of QESA. According to Ref[17], $\lambda=\frac{4}{3}$ is to select the best number of Grover iterations as quickly as possible. If $\lambda^{t-1}>\sqrt{N}$, then $\sqrt{N}$ will replace $\lambda^{t-1}$.

(3) **The comparison result**

The circuits of combination that $N=3$, $M=2$ and the initial state is $|\Psi\rangle$ based on Grover-Long algorithm (Fig 9(a)) and QESA (Fig 9(b, c)) are firstly compared. As aforementioned, the modified Grover-Long circuit requires fewer entanglement gates and does not require an ancilla qubit. And as for Grover-Long algorithm, number of iterations can be estimated in advance, thus it is deterministic.

Failure rate of two algorithms are then compared. In theory, $\varepsilon_{ESA}^{(6)} \approx \varepsilon_{GL}$, but in experiment, this relationship changes to $\varepsilon_{ESA_E}^{(3)} \approx \varepsilon_{GL_E}$. In most cases, the theoretical failure rates of Grover algorithm are more than 50% due to improper number of iterations. On the contrary, in the experiment, gate errors and readout errors can lead to chaos, which can counteract failure rate to around 50%. The specific gate errors and readout errors are attached in Appendix A.

To avoid redundancy, circuits of other combinations are not depicted, but the failure rates and measurement results of Grover-Long, as well as iteration number of QESA under approximate failure rate against Grover-Long are compared, as shown in Fig 10.

All the comparison of results indicates that our algorithm and circuits are more efficient than QESA and its circuits.

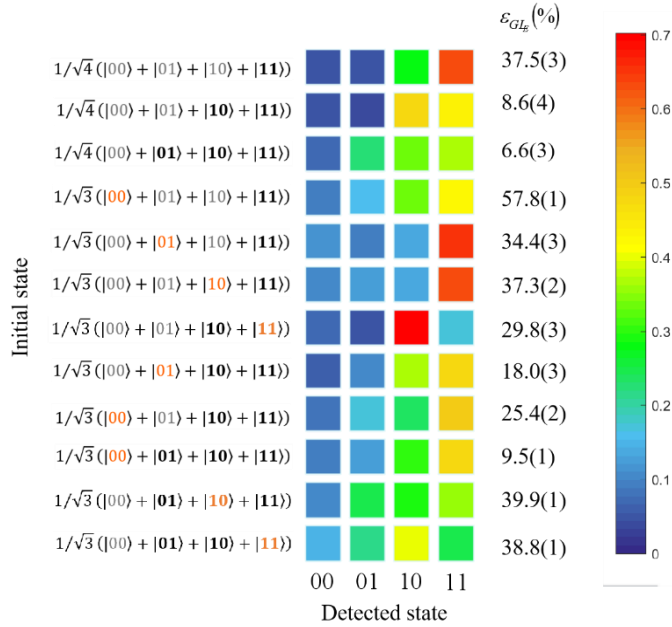

**Fig 10.** The comparison of experimental results from the execution of two algorithms performed on a 2-qubit database. The bold black state, the orange state denotes, the gray state and the bold orange state denote the marked states, the state with an amplitude of 0, the normal state, the wrong marked state, respectively. The gradient color map shows the probability of detecting each output state after applying Grover-Long algorithm. The experimental failure rate of Grover-Long algorithm $\varepsilon_{GL_E}$ is given and the number in parentheses indicates the number of QESA iterations when two algorithms have a similar experimental failure rate.

## 4.2 A 6-qubit contrast numerical simulation

To verify that our algorithm can efficiently solve minimum finding in a real dataset ($N = 36$ items excerpt from the passenger age of Titanic), we design and implement a 6-qubit numerical simulation experiment. The complete data of Titanic passenger age(excerpt) is attached in Appendix B. And the theoretical failure rates of the two algorithms are analyzed.

In simulation, firstly, the unsorted database was encoded into 6 qubits. The amplitude of an orthogonal basis state for a data value storing is $\frac{1}{\sqrt{N}}$. If there is no data value to store, the amplitude of an orthogonal basis state is 0. Then we started estimation, same as which in the experiment, the estimated number of solutions $\widetilde{M} = d_0 + 1$ and the estimated database size $\widetilde{N} = 2^n$, where n=6. We set the randomly selected reference value from the unsorted database is $d_0 = 47$. Then after applying Grover-Long algorithm, the measurement result $d_1$ could be obtained. If $d_1 \geq d_0$, the algorithm is thought to operate successfully. Owing to the error between estimation values and exact values of $M$, $N$, Grover-Long algorithm has a certain failure rate $\varepsilon_{GL}$. The complete circuit and failure rate of such specific case are shown in Fig 11.

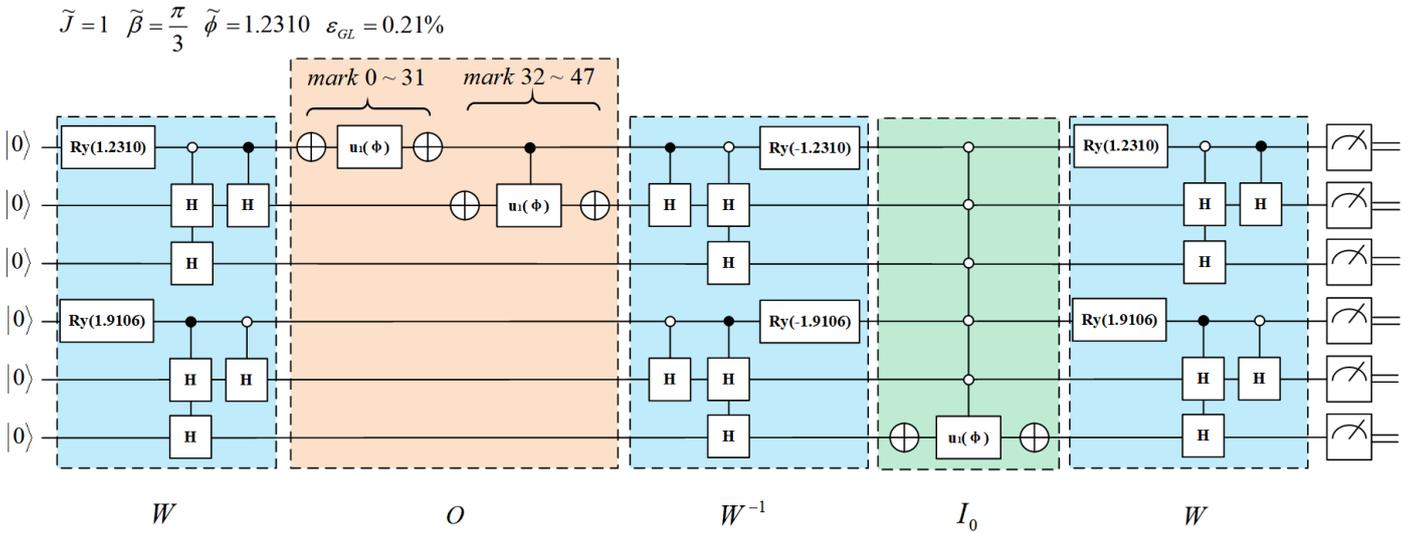

**Fig 11.** A 6-qubit circuit of Grover-Long algorithm with 36 items. The upper left corner of this picture shows some estimated parameters of Grover-Long algorithms.

Circuits construction and measurement under other $d_0$ is almost the same. Note that, $d_0$ might be equal to any data value in the unsorted database. Thus the error between the estimated value $\frac{\widetilde{M}}{\widetilde{N}}$ and the exact value $\frac{M}{N}$ varies as $d_0$ changing. Fig 10(a) showed the theoretical failure rate of Grover-Long algorithm when $M$ is different ($N$ and $\widetilde{N}$ are fixed). Similarly, Fig 10(b) shows he theoretical failure rate of QESA with different iterations.

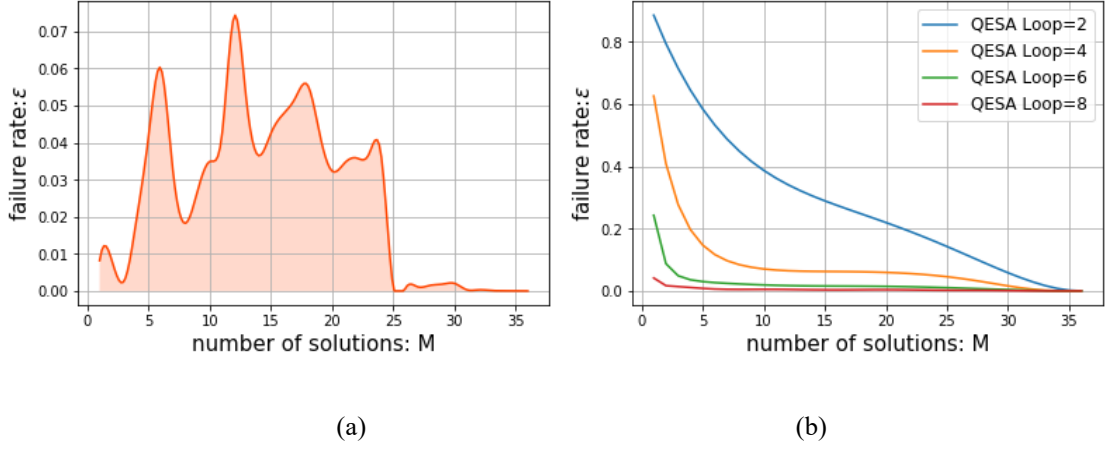

(a)                            (b)

**Fig 12**. The trend of the failure rate of two algorithms. (a) The Grover-Long algorithm. (b) QESA, where different color curves correspond to different number of QESA iterations.

As shown in Fig 12, we can draw a conclusion: our algorithm has a lower failure rate to solve a real issue. Unlike QESA probabilistically selecting the number of Grover iterations with multiple times, our algorithm has a determined number of iterations. Thus, we can avoid unnecessary the initial state preparation.

However, there are two questions about Grover-Long algorithm. First, the failure rate of Grover-Long algorithm will increase with data size growing, due to the unknown exact values of $M$, $N$. Second, is there a better way to estimate $M$, $N$ and further reduce the failure rate of Grover-Long algorithm, thereby improving the efficiency of QUMMSA. In the next section, we will analyze influencing factors of the failure rate.

## 5. Performance

We explain the performance of QUMMSA from two aspects: the failure rate and the complexity. Specifically, we propose two methods to reduce the failure rate and compare the complexity of DHA and our algorithm.

### 5.1 Failure rate

In this section, the failure rate can be contributed by two parts:

1. The failure rate of Grover-Long algorithm caused by the unknown number of solutions.

2. The failure rate in the main loop caused by the selection of the interrupt condition.

1. The failure rate of Grover-Long algorithm

For convenience, we only present the minimum value searching algorithm as an example. If we want to obtain the exact value of $\beta, J, \phi$, we must know $\frac{M}{N}$. But not all quantum states store a data value. When a random value $d_0$ is obtained, we do not know how many values are less than or equal to it. Namely, the ranking of $d_0$, which is denoted by $r_0$, is unknown in an unsorted database. We use estimated values $\widetilde{M}, \widetilde{N}$ to replace exact values $M, N$. Without any prior knowledge of the database, we assume that each orthonormal basis state stores a data value. Therefore, we assume that each orthonormal basis state stores a data value. Therefore, the estimated database has $\widetilde{N} = 2^n$ non-repeating values and the estimated number of marked states is $\widetilde{M} = d_0 + 1$. The distribution function of the estimated database is regarded as a uniform distribution, where the probability density function $\tilde{p}(x) = \frac{1}{\widetilde{N}}$. And the distribution function of the real database is unknown. It is important to quantify the impact of the gap between $\frac{\widetilde{M}}{\widetilde{N}}$ and $\frac{M}{N}$ on the failure rate $\varepsilon_{GL}$. We simulated the mapping between the above three, as shown in Fig 13.

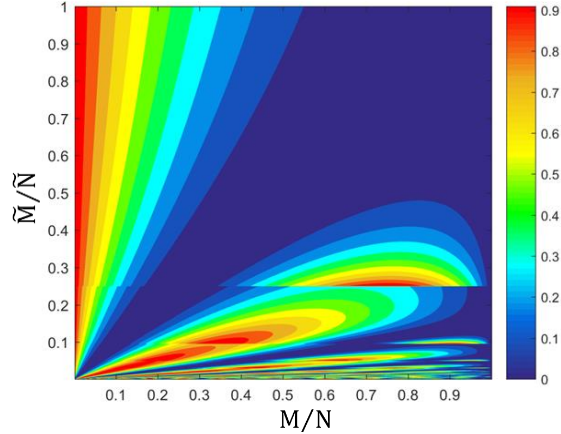

**Fig 13.** A contour map for the mapping between $\frac{\widetilde{M}}{\widetilde{N}}, \frac{M}{N}$ and the failure rate $\varepsilon_{GL}$, where X label is $\frac{M}{N}$, Y label is $\frac{\widetilde{M}}{\widetilde{N}}$. Ten colors correspond to different levels of the failure rate.

The cumulative distribution function of the estimated database in $[0, d_0]$ is described as:

$$\widetilde{P}(x) = \int_0^{d_0} \tilde{p}(x) dx = \frac{\widetilde{M}}{\widetilde{N}} \tag{13}$$

Similarly, the cumulative distribution function of the real database $P(x)$ is $\frac{M}{N}$. Therefore, we can draw a conclusion: if the real database obeys uniform distribution in $[0, 2^n - 1]$, then $\frac{\widetilde{M}}{\widetilde{N}} : \frac{M}{N} \approx 1$. In other words, the failure rate of the area near the diagonal is close to 0, as shown in Fig 13. However, most databases don't satisfy this restriction. We propose a method to reduce the failure rate.

**How to reduce the failure rate?**

We consider a new method that uses the sampling distribution function to replace the overall data distribution. When the overall data exceeds millions, the minimum sample size is independent of the number of overall data and is related to the confidence level, acceptable error, degree of dispersion between samples. The minimum sample size is given [37]:

$$h = \frac{Z^2 \sigma^2}{E^2} \tag{14}$$

Where $Z$ is Z-statistic; $\sigma^2$ is variance; $E$ is an acceptable error. The minimum sample sizes with different confidence levels, acceptable errors are shown in Appendix D. Note that, the estimated values $\widetilde{M}$, $\widetilde{N}$ is unimportant. As long as $\widetilde{P}(x) \approx P(x)$, we can calculate the parameters of Grover-Long algorithm, accurately. Taking the number of Grover-Long iterations as a baseline, we compare the failure rate of QESA and Grover-Long algorithm when we use this new method. The result is shown in Fig 14.

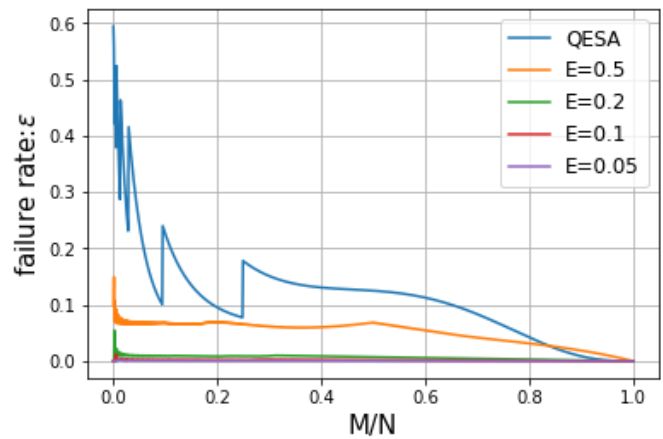

**Fig 14.** Comparison of the failure rate for QESA and Grover-Long algorithm. The blue curve donates the failure rate of QESA. Other curves donate the failure rate of Grover-Long algorithm with different acceptable error.

2. The failure rate of main loop

This part of the failure rate is due to the fact that we don't know when to end the main loop. We can obtain $d_1$, less than or equal to $d_0$. But we don't know $d_1$ whether is the minimum value. If we finish the main loop too early, then we only obtain a minor value rather than the minimum value. In contrast, we will waste unnecessary computing resources. Therefore, it is urgent to find a suitable opportunity to end the main loop.

**How to reduce the failure rate?**

If Grover-Long algorithm is executed successfully and we obtain $d_1$. the possibility of $d_1 = d_0$ will be $\frac{1}{r_0}$, where $r_0$ is the rank of $d_0$ in the unsorted database, since the output of Grover-Long algorithm has an equal possibility. The probability that $d_1 = d_0$ consecutive occurrences of $c$ times is $\left(\frac{1}{r_0}\right)^c$. Namely, the possibility of $d_0$ that is not the minimum value is $\left(\frac{1}{r_0}\right)^c$ (the worst case is $r_0 = 2$). Therefore, we only add few main loops and can find the minimum value close to 100% successful probability. This improvement has been applied in QUMMSA pseudocode in Section 2.

## 5.2 Complexity

By convention, we reaffirm that the complexity of QUMMSA is primarily made up a total number of Grover-Long iterations and the initial state preparation. Other steps of QUMMSA are not counted.

One main loop possesses $J$ Grover-Long iterations. $J$ can be described as formula (15):

$$J = \max\left[\text{floor}\left(\frac{\frac{\pi}{2} - \arcsin\left(\sqrt{\frac{M}{N}}\right)}{\arcsin\left(\sqrt{\frac{M}{N}}\right)}\right), 1, \text{ceil}\left(\frac{\pi - 6\arcsin\left(\sqrt{\frac{M}{N}}\right)}{4\arcsin\left(\sqrt{\frac{M}{N}}\right)}\right)\right] \quad (15)$$

Where floor($\cdot$) is rounding down to an integer; ceil($\cdot$) is rounding up to an integer.

Considering an infinity database:

$$\lim_{\sqrt{\frac{M}{N}} \to 0} \arcsin\left(\sqrt{\frac{M}{N}}\right) \approx \sqrt{\frac{M}{N}} \quad (16)$$

The complexity of Grover-Long algorithm is simplified to formula (17).

$$J = \max\left[\frac{\pi}{2} \times \sqrt{\frac{N}{M}}, \quad \frac{\pi}{4} \times \sqrt{\frac{N}{M}} - \frac{3}{2}\right] = \frac{\pi}{2}\sqrt{\frac{N}{M}} \tag{17}$$

The total of Grover-Long iterations can be described as formula (18), when QUMMSA is executed completely.

$$R_G = \sum_{k=0}^{K-1} J_k = \frac{\pi}{2}\sum_{k=0}^{K-1}\sqrt{\frac{N}{M_k}} \tag{18}$$

Where $K$ denotes the total number of main loops; $M_k$ denotes the number of marked states of the $k$-th main loop. Since the output of Grover-Long algorithm has an equal possibility, the $k$-th main loop's the number of marked quantum states is nearly double of the $(k+1)$-th main loop's. Thus, the trend of $M_k$ is regarded as a geometric progression. The complexity of all Grover-Long iterations can be described as formula (19).

$$R_G = \frac{\pi}{2}\left(\sqrt{\frac{N}{M_0}} + \sqrt{\frac{2N}{M_0}} + \cdots + \sqrt{\frac{M_0 N}{M_0}}\right) = \frac{\pi}{2}\frac{\sqrt{\frac{N}{M_0}} \times (1 - \sqrt{2M_0})}{1 - \sqrt{2}} = \frac{\pi}{2}(\sqrt{2}+1)\left(\sqrt{2N} - \sqrt{\frac{N}{M_0}}\right) \tag{19}$$

Next, we consider the complexity of the initial state preparation, It needs to be executed approximately $\log_2 N$ times. Meanwhile, it takes $\log_2 N$ time steps when it is executed once.

$$R_{init} = \log_2^2 N \tag{20}$$

$M_0$ is the number of marked quantum states in the first main loop. Usually, $M_0 \approx \frac{1}{2}N$. Considering the failure rate and the interrupt condition, we obtain the final complexity.

$$R = \frac{\frac{\pi}{2}(2 + \sqrt{2} + c)\sqrt{N} + (\log_2 N + c)\log_2 N}{1 - \varepsilon} \tag{21}$$

Finally, we compare two algorithms in the same condition. We set $\varepsilon = 0.1$ and calculate the complexity of two algorithms with different database sizes, as shown in Fig 15. Two curves show that our algorithm has a greater advantage with the increasing of database size, since QUMMSA only requires to prepare a smaller number of initial states (QUMMSA and DHA approximately require $\log_2 N$ and $\log_2^2 N$ times, respectively) and fewer oracle calls. Meanwhile, we note that the lower bound of the complexity for the initial state preparation is $\log_2 N$. If the initial state becomes more complex, QUMMSA will have a greater advantage.

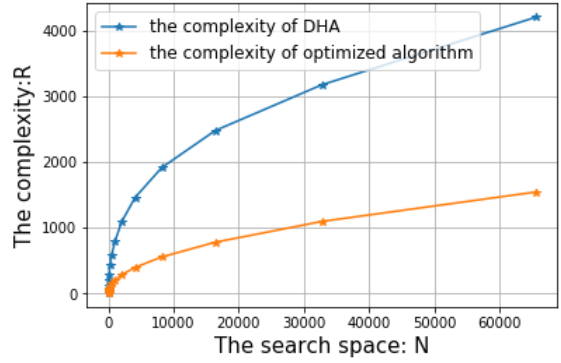

**Fig 15.** Complexity comparison for two algorithms.

## 6. Conclusions

In summary, we demonstrate the advantage of the quantum algorithm for finding maximum or minimum value to alleviate some of the challenges brought by the rapidly increasing amount of data. Based on Grover-Long algorithm, we proposed QUMMSA, which is an improved version of DHA. Compared with DHA, it has close to 100% successful probability and greater advantage in complexity with the increasing database size. Furthermore, we provide the corresponding general quantum circuits and three equivalent simplified principles which reduce gate complexity of constructing oracles. The optimized circuits are easy to implement on any general-purpose quantum computer. Meanwhile, the general design method for circuits can be integrated into quantum software. we demonstrate the advantage of our algorithm through a group 2-qubit contrast experiment which is executed on a superconducting processor and a real issue that is numerical simulated.

In addition to the computational tasks we show in this paper, the algorithm can be a subroutine in other quantum algorithms that need to find a maximum or minimum value. We hope that the theoretical and experimental results we present here well push further research and motivate innovations of other mathematical models. The paradigm combing classical steps and quantum steps may work as an efficient solution in the era of big data.

## Authorship Contributions

Yanhu Chen provided the original idea and performed quantum simulations. Shijie Wei, Jian Wu, Hongxiang Guo made constructive comments. and Shijie Wei, Xiong Gao, Cen Wang modified the manuscript and discussed the detail. All authors contributed to manuscript writing and discussions about the results.

# Appendix A

In this section, we present some parameters of IBM quantum superconducting processor. It is shown in Table 1.

Table 1 Some parameters of IBM quantum superconducting processor

| Qubit Label | Q0 | Q1 | Q2 |
|---|---|---|---|
| Frequency(GHz) | 5.250 | 5.30 | 5.34 |
| T1(μs) | 58.90 | 25.40 | 42.20 |
| T2(μs) | 22.20 | 8.90 | 54.70 |
| Single Qubit Gate Error($10^{-3}$) | 0.77 | 6.36 | 1.20 |
| Readout Error($10^{-2}$) | 5.00 | 7.50 | 2.90 |
| Multi Qubit Gate Error($10^{-2}$) | | CX1_0 2.61 | CX2_0 2.40 |
| | | | CX2_1 5.42 |

# Appendix B

This section shows data source of the second demo. Complete data can be obtained on Kaggle website (https://www.kaggle.com/c/titanic/data). The complete data we use is listed in Table 2.

**Table 2** Titanic passengers age(excerpt)

| Name | Age | After Encoding | Name | Age | After Encoding | Name | Age | After Encoding |
|---|---|---|---|---|---|---|---|---|
| Gee, Mr. Arthur H | 47 | 101111 | Harris, Mr. George | 62 | 111110 | Elias, Mr. Tannous | 15 | 001111 |
| Rice, Master. Eric | 7 | 000111 | Coxon, Mr. Daniel | 59 | 111011 | Danoff, Mr. Yoto | 27 | 011011 |
| Skoog, Miss. Mabel | 9 | 001001 | Rugg, Miss. Emily | 21 | 010101 | Nicola-Yarred, Master. Elias | 12 | 001100 |
| Ekstrom, Mr. Johan | 45 | 101101 | Burke, Mr. Jeremiah | 19 | 010011 | Ling, Mr. Lee | 28 | 011100 |
| McKane, Mr. Peter David | 46 | 101110 | Christmann, Mr. Emil | 29 | 011101 | West, Miss. Constance Mirium | 5 | 000101 |
| Panula, Master. Eino Viljami | 1 | 000001 | Turkula, Mrs. (Hedwig) | 63 | 111111 | Saad, Mr. Khalil | 25 | 011001 |
| Hansen, Mr. Claus Peter | 41 | 101001 | Cherry, Miss. Gladys | 30 | 011110 | Skoog, Master. Harald | 4 | 000100 |
| Hold, Mr. Stephen | 44 | 101100 | Hampe, Mr. Leon | 20 | 010100 | Cook, Mr. Jacob | 43 | 101011 |
| Mack, Mrs. (Mary) | 57 | 111001 | Moor, Master. Meier | 6 | 000110 | Ayoub, Miss. Banoura | 13 | 001101 |
| Sutton, Mr. Frederick | 61 | 111101 | Weir, Col. John | 60 | 111100 | Calic, Mr. Jovo | 17 | 010001 |
| Waelens, Mr. Achille | 22 | 010110 | Zabour, Miss. Hileni | 14 | 001110 | Palsson, Miss. Stina Viola | 3 | 000011 |
| Newell, Miss. Madeleine | 31 | 011111 | Hassan, Mr. Houssein G N | 11 | 001011 | Pain, Dr. Alfred | 23 | 010111 |

# Appendix C

In this section, we given the failure rate of Grover-Long algorithm and QESA, when the number of solutions is unknown in theory.

First, we present a proof of Grover-Long algorithm's failure rate $\varepsilon_{GL}$, when $M$ is unknown. The performance is shown in Fig 13.

The initial quantum state is expressed as:

$$|\Psi\rangle = \left(\sqrt{\frac{M}{N}}|\Psi_{\text{good}}\rangle + \sqrt{\frac{N-M}{N}}|\Psi_{\text{bad}}\rangle\right) \quad \text{(C1)}$$

Where $|\Psi_{\text{good}}\rangle$ includes $M$ solutions. $|\Psi_{\text{bad}}\rangle$ includes $N-M$ non-solutions. Each quantum state store a data value and their amplitude is expressed as $\alpha_{good}^{(0)} = \alpha_{bad}^{(0)} = \sqrt{\frac{1}{N}}$, in the initial state. Otherwise, the amplitude will be 0, if no data value is stored.

Grover-Long algorithm is divided into 4 steps. The first step is the oracle operator. It makes solutions receive a phase shift $\phi$:

$$O|\Psi\rangle = \begin{bmatrix} e^{i\phi} & \\ & 1 \end{bmatrix}(|\Psi_{good}\rangle + |\Psi_{bad}\rangle) \tag{C2}$$

The step (2)(3)(4) can be expressed as:

$$-W\left((e^{i\phi} - 1)|0\rangle\langle 0| - I\right)W^{-1} = -(e^{i\phi} - 1)|\Psi\rangle\langle\Psi| - I \tag{C3}$$

Where $W$ is a unitary operator to prepare the initial state.

Therefore, Grover-Long algorithm can be expressed as:

$$G|\Psi\rangle\langle\Psi| = \left(-(e^{i\phi} - 1)|\Psi\rangle\langle\Psi| - I\right)(e^{i\phi}|\Psi_{good}\rangle + |\Psi_{bad}\rangle) \tag{C4}$$

After $G$ operator, we obtain two results. For solutions, each quantum state's amplitude will be expressed as:

$$\alpha_{good}^{(j)} = -\alpha_{good}^{(j-1)} \times \left(1 + \frac{e^{i\tilde{\phi}} - 1}{N}\right) - \alpha_{good}^{(j-1)}(M - 1) \times \frac{e^{i\tilde{\phi}} - 1}{N} - \alpha_{bad}^{(j-1)} \times (N - M) \times \frac{e^{i\tilde{\phi}} - 1}{N} \tag{C5}$$

For non-solutions, each quantum state's amplitude will be expressed as:

$$\alpha_{bad}^{(j)} = -\alpha_{bad}^{(j-1)} \times \left(1 + \frac{e^{i\tilde{\phi}} - 1}{N}\right) - \alpha_{good}^{(j-1)} \times M \times \frac{e^{i\tilde{\phi}} - 1}{N} - \alpha_{bad}^{(j-1)} \times (N - M - 1) \times \frac{e^{i\tilde{\phi}} - 1}{N} \tag{C6}$$

Where $j$ is the current number of iterations and $j \in [1, \tilde{J}]$. The Grover-Long estimated parameters $\tilde{J}$, $\tilde{\phi}$ are calculated by formula(3,4).

The failure rate is expressed as:

$$\varepsilon_{GL} = 1 - M \times \text{abs}\left(\alpha_{good}^{(J)}\right)^2 \tag{C7}$$

We mark all quantum states as less than or equal to $d_0$, regardless of whether the quantum state stores a data value. Thus the estimated number of solutions $\tilde{M} \geq M$. Marking quantum states of 0 amplitude does not affect the iterative process. If the amplitude of a quantum state is 0, the amplitude will be still 0 after the amplitude amplification.

Second, we present proof of QESA's failure rate $\varepsilon_{ESA}$. The performance is shown in the blue curve of Fig 14. Due to the unknown $M$, different number of Grover iterations is selected in different possibility in once QESA iteration. Specially, Ref[17] set a parameter $\lambda \in (1, \frac{4}{3}]$. The number of Grover iterations $v$ is a random number which is selected from $[0, \lambda^{t-1})$ and rounds down where $t$ is the current number of QESA iteration.

In the first QESA iteration.

$$\varepsilon_{ESA}^{(1)} = 1 - \frac{M}{N} \tag{C8}$$

If the algorithm doesn't find a correct solution, it will run forever. Meanwhile, $\varepsilon_{ESA}^{(t)}$ is decreased with the increase of $t$.

$$\varepsilon_{ESA}^{(t)} = \varepsilon_{ESA}^{(t-1)} \times \left\{ \lambda^{-t+1}\left(1 - \frac{M}{N}\right) + \left[\sum_{\nu=1}^{\text{floor}(\lambda^{t-1})-1} \lambda^{-t+1}\cos^2\left((2\nu+1) \times \arcsin\sqrt{\frac{M}{N}}\right)\right] + [\lambda^{t-1} - \text{floor}(\lambda^{t-1})]\cos^2\left((2 \times \text{floor}(\lambda^{t-1}) + 1) \times \arcsin\sqrt{\frac{M}{N}}\right) \right\} \tag{C9}$$

If $\lambda^{t-1} > \sqrt{N}$, then $\sqrt{N}$ will replace $\lambda^{t-1}$.

Through these proofs, we can quickly simulate the failure rate of two algorithms and avoid massive unitary matrices transforms.

## Appendix D

In this section, we present the minimum sample size in different confidence levels, acceptable errors as shown in Table 3.

**Table 3** The minimum sample size

| C \ E | 50% | 75% | 80% | 85% | 95% | 99% | 99.90% |
|---|---|---|---|---|---|---|---|
| 0.01 | 1140 | 3307 | 4096 | 5184 | 9604 | 16590 | 19741 |
| 0.03 | 127 | 358 | 456 | 576 | 1068 | 1844 | 2194 |
| 0.05 | 46 | 133 | 164 | 208 | 385 | 664 | 790 |
| 0.1 | 12 | 34 | 41 | 52 | 97 | 166 | 198 |
| 0.15 | 6 | 15 | 19 | 24 | 43 | 74 | 88 |
| 0.2 | 3 | 9 | 11 | 13 | 25 | 42 | 50 |

Where $C$ is confidence level; $E$ is an acceptable error.